# Stable soliton complexes and azimuthal switching in modulated Bessel optical lattices

Yaroslav V. Kartashov,[1,2] Alexey A. Egorov,[2] Victor A. Vysloukh,[3] Lluis Torner[1]

[1]*ICFO-Institut de Ciencies Fotoniques and Department of Signal Theory and Communications, Universitat Politecnica de Catalunya, 08034 Barcelona, Spain*
[2]*Physics Department, M. V. Lomonosov Moscow State University, 119899, Moscow, Russia*
[3]*Departamento de Fisica y Matematicas, Universidad de las Americas – Puebla, Santa Catarina Martir, Cholula, 72820, Puebla, Mexico*

We address azimuthally modulated Bessel optical lattices imprinted in focusing cubic Kerr-type nonlinear media, and reveal that such lattices support different types of stable solitons, whose complexity increases with growth of lattice order. We reveal that the azimuthally modulated lattices cause single solitons launched tangentially to the guiding rings to jump along consecutive sites of the optical lattice. The position of the output channel can be varied by small changes of the launching angle.

Weak transverse modulation of refractive index of nonlinear medium creates inhomogeneities that are capable to capture and hold optical radiation. In the simplest case of periodic modulation of refractive index, the waveguide arrays enable the existence of discrete solitons formed upon the competition of discrete diffraction and nonlinearity [1]. Lattices with tunable strength introduce new opportunities for soliton control [2]. It was recently shown that such tunable lattices can be induced optically in photorefractive medium [3-9], a possibility that enables controlling both the lattice period and the refractive index modulation depth. The basic properties of one- and two-dimensional solitons (including dipole-mode ones and soliton complexes) supported by harmonic lattices are now well established [2-11].

In addition to the concept of *tunable discreteness*, the landmark idea of all-optical lattice generation with nondiffracting fields opens broad prospects for creation and exploration of lattices with different symmetry. For example, we recently put forward the properties of solitons supported by *radially symmetric optical lattices*, induced by non-diffracting Bessel beams [12]. Such beams can be created by illuminating a conical shaped optical element, called an axicon, with a Gaussian beam, or by using a narrow illuminated annular slit that is placed in the focal plane of a focusing lens [13]. More elaborated holographic techniques can be used to produce the *higher-order azimuthally modulated nondiffracting beams and lattices*. In this paper we introduce the idea of solitons supported by such type of lattices, and show that they support soliton complexes that intuitively can be viewed as nonlinear combinations of several lowest-order solitons. We also show that solitons can be set into azimuthal rotation when launched tangentially to the main guiding ring at appropriate angles. Such rotation is accompanied by small radiation that finally leads to the controllable trapping of the solitons in one lattice ring.

We consider propagation of optical radiation along the $z$ axis of a bulk focusing cubic medium with transverse modulation of linear refractive index described by the nonlinear Schrödinger equation for dimensionless complex field amplitude $q$:

$$i\frac{\partial q}{\partial \xi} = -\frac{1}{2}\left(\frac{\partial^2 q}{\partial \eta^2} + \frac{\partial^2 q}{\partial \zeta^2}\right) - q|q|^2 - pR(\eta,\zeta)q. \qquad (1)$$

Here the longitudinal $\xi$ and transverse $\eta,\zeta$ coordinates are scaled to the diffraction length and input beam width, respectively. The parameter $p$ describes the lattice depth. The profile of the modulated lattice is given by $R(\eta,\zeta) = J_n^2[(2b_{\text{lin}})^{1/2}r]\cos^2(n\phi)$, where $r = (\eta^2 + \zeta^2)^{1/2}$ is the radius, $\phi$ is the azimuth angle, and the parameter $b_{\text{lin}}$ defines the transverse lattice scale. Note that a function $q(\eta,\zeta,\xi) = J_n[(2b_{\text{lin}})^{1/2}r]\cos(n\phi)\exp(-ib_{\text{lin}}\xi)$ describes a higher-order azimuthally modulated Bessel beam creating the lattice and is an exact solution of the linear homogeneous Schrodinger equation. We assume that the lattice profile mimics the intensity profile of the nondiffracting beam, as it occurs in photorefractive crystals. Due to a specific field distribution in higher-order Bessel beams the depth of azimuthal refractive index modulation in the lattice is 100%. Note that with several incoherent Bessel beams of different intensities and orders it is also possible to produce lattices that are weakly modulated in azimuthal direction. It is assumed that the depth of the refractive index modulation is small compared with unperturbed refractive index and is of the order of nonlinear contribution to refractive index. Optical induction of lattices in photorefractive crystals is possible because of the large anisotropy of their nonlinear response. While linear anisotropy has almost no effect on propagation of the lattice-creating Bessel beam (because of the small relative difference $(\chi_{xx} - \chi_{yy})/\chi_{xx,yy} \sim 10^{-3}$ between elements of susceptibility tensor), only the nonlinear anisotropy breaks rotational symmetry and affects the properties of solitons supported by Bessel lattices. We checked by solving the full system of material equations for photorefractive crystals that main results (e.g. possibility of azimuthal switching and existence of stable soliton complexes) obtained with the model (1) remain valid in the presence of anisotropy of nonlinear response. Nevertheless, here we use model (1) since it also holds for trapped Bose-Einstein condensates. Typical profiles of modulated Bessel lattices with $n = 1,2$ are shown in Fig. 1. The local lattice maxima situated closer to the lattice center are more pronounced than others and form a ring of guiding channels that will be referred to as the main guiding ring of the lattice. The number of guiding channels in the main ring is given by $2n$, while the radius of the main ring increases with growth of the lattice order. Eq. (1) admits conserved quantities, including the power or energy flow $U = \int_{-\infty}^{\infty} \int_{-\infty}^{\infty} |q|^2 \, d\eta d\zeta$.

We searched for solutions of Eq. (1) in the form $q(\eta, \zeta, \xi) = w(\eta, \zeta) \exp(ib\xi)$, where $w(\eta, \zeta)$ is a real function and $b$ is a real propagation constant. Soliton families are defined by the propagation constant $b$, the parameter $b_{\text{lin}}$, the lattice order $n$, and the lattice depth $p$. Since scaling transformations $q(\eta, \zeta, \xi, p) \to \chi q(\chi\eta, \chi\zeta, \chi^2\xi, \chi^2 p)$ can be used to obtain various families of lattice solitons from a given one, here we set the transverse scale in such way that $b_{\text{lin}} = 2$, and vary $b$, $p$, and $n$. To elucidate the linear stability of solitons we searched for perturbed solutions in the form $q(\eta, \zeta, \xi) = [w(\eta, \zeta) + u(\eta, \zeta, \xi) + iv(\eta, \zeta, \xi)]\exp(ib\xi)$, with $u$ and $v$ being the real and imaginary parts of the perturbations which can grow upon propagation with a complex rate $\delta$. A standard linearization procedure for Eq. (1) yields a system of coupled

Schrödinger-type equations for perturbation components $u,v$ that we solved numerically, in order to find perturbation profiles and growth rate.

One-dimensional and two-dimensional soliton configurations can be stable only when the field changes sign between neighboring channels. Since the highest refractive index modulation occurs for the main guiding ring of the modulated lattice, it is natural to expect that the main guiding ring of a $n$-th order Bessel lattice can support stable soliton complexes formed by $2n$ out-of-phase bright spots. The properties of simplest soliton complexes or dipole solitons supported by the first-order lattice are summarized in Fig. 2. The typical profile of dipole solitons, found with a standard relaxation method, is shown in Fig. 2(a). Such soliton can be intuitively viewed as nonlinear combination of two out-of-phase lowest orders solitons supported by two guiding sites of main ring of Bessel lattice. The lattice compensates the repulsive interaction between out-of-phase solitons and makes possible their propagation as a single entity. Energy flow $U$ of the dipole soliton is a nonmonotonic function of propagation constant (Fig. 2(b)). At high energy flows when $b \to \infty$ two spots forming dipole become narrow and almost do not interact. At small lattice depth $p \leq p_{\mathrm{cr}}$, where $p_{\mathrm{cr}} \approx 3$, dipole soliton drastically broadens with diminishing of the propagation constant and, as propagation constant approaches termination point, soliton ceases to exist. At $p > p_{\mathrm{cr}}$ energy flow of dipole soliton vanishes in the cutoff, while its azimuthal width changes slightly. This behavior corresponds to discontinuity in the cutoff versus lattice depth curve (Fig. 2(c)). A comprehensive study of the linear stability of the dipole solitons supported by Bessel lattices with moderate depth revealed the existence of an instability domain located close to the cutoff on the propagation constant (Fig. 2(d)). The corresponding instability is of an oscillatory type, and typically $\mathrm{Re}(\delta) \ll \mathrm{Im}(\delta)$. Both the width of the instability domain and the maximal real part of complex growth rate were found to decrease with growth of the lattice depth. For deep enough lattices dipole solitons become free from instabilities in the entire domain of their existence.

Properties of quadrupole solitons supported by second-order lattice are summarized in Fig. 3. Quadrupole solitons can be viewed as nonlinear superposition of four out-of-phase bright spots. Their properties are qualitatively similar to that of dipole solitons. There exist lower cutoff on propagation constant that is a nonmonotonic function of the lattice depth with a discontinuity at $p_{\mathrm{cr}} \approx 7$ (Fig. 3(c)). A careful linear stability analysis revealed that the structure of instability domain for quadrupole solitons (located near the propagation constant cutoff) is rather complex, with separate "stability windows", and both oscillatory and exponential instabilities may take place for such solitons. Thus Fig. 3(d) shows the widest instability band for quadrupole soliton at $p=16$. At moderate lattice depth instability vanishes above a certain propagation constant threshold. The instability domain was found to shrink for deep lattices, and quadrupole solitons were found to become stable in the entire domain of their existence.

To stress that the concept of stable soliton complexes can be generalized to higher-order structures, we have studied higher-order lattices with $n$ up to 10. All of them can support soliton complexes, whose properties are similar to that of dipole and quadrupole solitons (see Fig. 4 with examples of stable higher-order soliton complexes). Note that in homogenous media soliton complexes (or clusters) tend to self-destroy through expansion or coalescence [14]. The instability may be reduced by the presence of competing nonlinearities, but even then the complexes exist as metastable objects [15].

Another intriguing opportunity afforded by azimuthally modulated Bessel lattices is that a single soliton initially located in one of the guiding sites of the main lattice ring and launched tangentially to the ring starts to travel along consecutive guiding sites of the ring, so that it can even return to the input site. In pure cubic medium soliton is strongly perturbed when it leaves the guiding site, making the above process probably difficult to observe in practice. However, even small nonlinearity saturation makes the process very robust. To illustrate this point we included into the model Eq. (1) small nonlinearity saturation by rewriting the nonlinear term as $-q|q|^2/(1+S|q|^2)$, where $S \ll 1$. In such case laser beam does not broaden in between guiding sites, and thus it is allowed to jump from one site to another. The soliton beam was set in motion by imposing on it an initial phase tilt $\exp(i\alpha_\eta \eta + i\alpha_\zeta \zeta)$. Here we consider the situation when the soliton beam is initially located in the outermost left guiding site of the main ring, and $\alpha_\eta = 0$. The soliton leaves the guiding site when $\alpha_\zeta$ exceeds a certain critical value, and starts to travel along the guiding ring if $\alpha_\zeta$ is not too high. Since solitons have to overcome a potential barrier when passing between neighboring sites, they radiate a small fraction of energy. In the presence of radiation soliton can be trapped in different guiding sites of the main ring and position/number of output site can be controlled easily by changing the launching angle or soliton energy flow (Fig. 5). Thus a higher incident angle is typically required to achieve trapping of soliton with higher energy flow into the desired guiding site. The potential of the effect to implement controllable azimuthal soliton switching is clearly visible.

Finally, we note that the azimuthal modulation of the lattice remarkably affects interactions experienced by solitons located in different guiding channels. When solitons carry identical energies formation of even or dipole soliton is possible. However, we found that when solitons carry different energies, they may fuse into a single soliton, independently of the phase difference between input solitons. Fig. 6 shows the input and output field distributions for different energy flows of a control soliton, in the case when the control and the input solitons are out-of-phase. If the energy flow $U_\mathrm{c}$ of the control soliton considerably exceeds that of input soliton all energy is concentrated in the site where control soliton was located (Figs. 6(c), 6(d)). When energy flows are comparable the output soliton can be located in the same channel as input one (Figs. 6(a) and 6(b)).

In conclusion we showed that azimuthally modulated Bessel optical lattices support soliton complexes that can be made stable in wide regions of their existence domain by varying the lattice strength. We also showed that single solitons launched tangentially to main guiding ring of the lattice can be trapped by its different guiding sites depending on the input angle and energy flow. Note that optically induced modulated Bessel lattices may find analogy with microstructured or photonic crystals.

This work has been partially supported by the Generalitat de Catalunya, and by the Spanish Government through grant BFM2002-2861.

# Figure captions

Figure 1.  Azimuthally modulated Bessel optical lattices of first two orders. All quantities are plotted in arbitrary dimensionless units.

Figure 2.  (a) Soliton supported by first-order Bessel lattice corresponding to point marked by circle in dispersion diagram (b). (c) Cutoff on propagation constant versus lattice depth. (d) Real part of perturbation growth rate versus propagation constant at $p=10$. All quantities are plotted in arbitrary dimensionless units.

Figure 3.  (a) Soliton supported by second-order Bessel lattice corresponding to point marked by circle in the dispersion diagram (b). (c) Cutoff on propagation constant versus lattice depth. (d) Real part of perturbation growth rate versus propagation constant at $p=16$. All quantities are plotted in arbitrary dimensionless units.

Figure 4.  (a) Stable soliton complex supported by third-order Bessel lattice at $b=3$ and $p=20$. (b) Stable soliton complex supported by sixth-order Bessel lattice at $b=5$ and $p=40$. All quantities are plotted in arbitrary dimensionless units.

Figure 5.  Azimuthal switching of soliton beam to second (a) and fourth (b) main guiding channels of third-order Bessel lattice imprinted in a cubic medium with nonlinearity saturation, for input angles $\alpha_\zeta = 0.49$ and $\alpha_\zeta = 0.626$ at $p=2$. Energy flow of input soliton $U_{\text{in}} = 8.26$. (c) and (d) show switching to third and sixth channels of sixth-order lattice for $\alpha_\zeta = 0.8$ and $\alpha_\zeta = 0.93$ at $p=5$. Energy flow of input soliton in (c) and (d) $U_{\text{in}} = 8.61$. Input and output intensity distributions are superimposed onto each other. Arrows show direction of input soliton motion and labels $S_{\text{in}}, S_{\text{out}}$ denote input and output soliton positions. Saturation parameter $S=0.1$. All quantities are plotted in arbitrary dimensionless units.

Figure 6. Controllable fusion of a soliton and control soliton beam launched into fifth guiding channel of third-order Bessel lattice imprinted in a cubic medium with nonlinearity saturation. Input (a) and output (b) intensity distributions for control beam energy $U_c = 10.64$. Input (c) and output (d) intensity distributions for $U_c = 20.29$. The input soliton energy is $U_{in} = 8.26$ and it is initially located in the zero channel. Labels $S_{in}, S_{out}, S_c$ denote input, output, and control beam positions. Saturation parameter $S = 0.1$, lattice depth $p = 2$. All quantities are plotted in arbitrary dimensionless units.

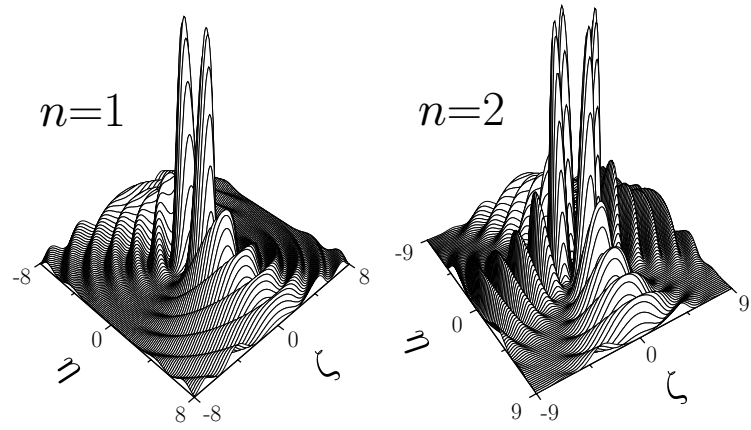

Figure 1.   Azimuthally modulated Bessel optical lattices of first two orders. All quantities are plotted in arbitrary dimensionless units.

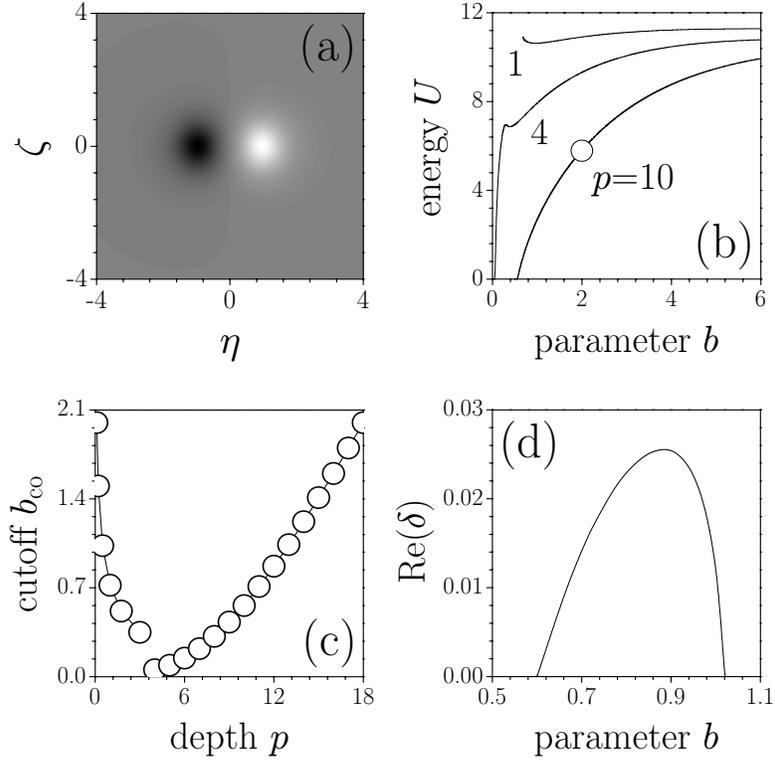

Figure 2. (a) Soliton supported by first-order Bessel lattice corresponding to point marked by circle in dispersion diagram (b). (c) Cutoff on propagation constant versus lattice depth. (d) Real part of perturbation growth rate versus propagation constant at $p = 10$. All quantities are plotted in arbitrary dimensionless units.

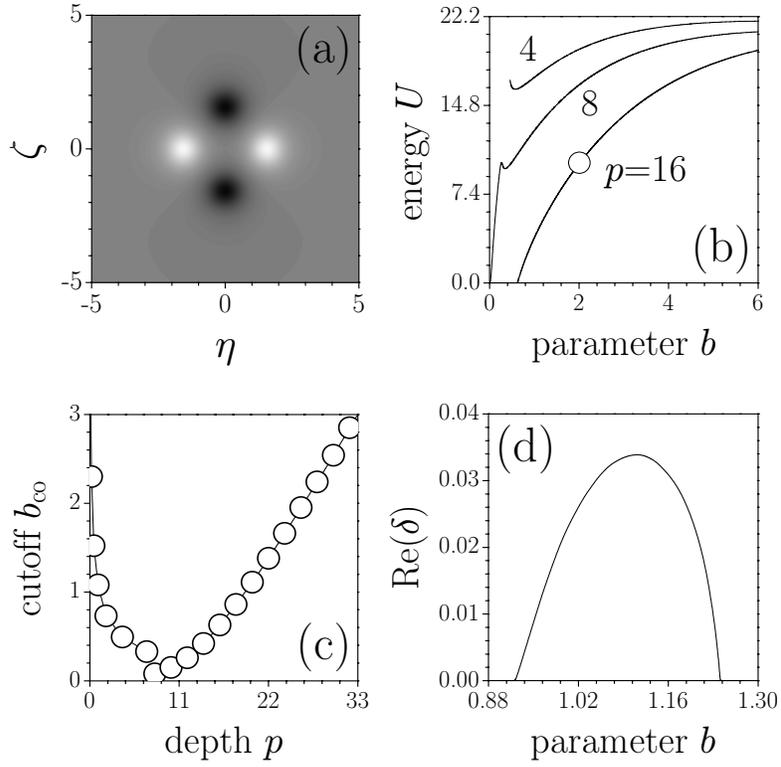

Figure 3. (a) Soliton supported by second-order Bessel lattice corresponding to point marked by circle in the dispersion diagram (b). (c) Cutoff on propagation constant versus lattice depth. (d) Real part of perturbation growth rate versus propagation constant at $p=16$. All quantities are plotted in arbitrary dimensionless units.

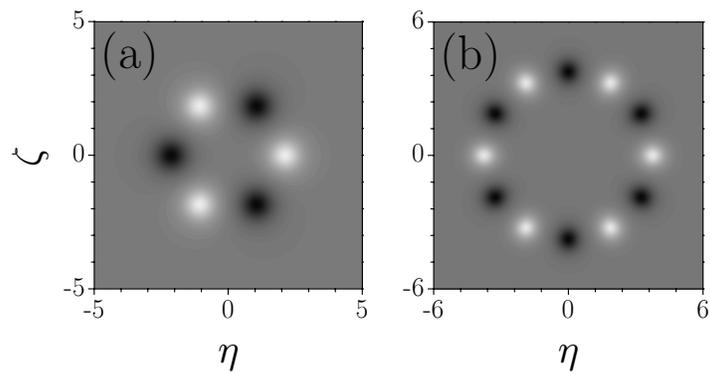

Figure 4. (a) Stable soliton complex supported by third-order Bessel lattice at $b=3$ and $p=20$. (b) Stable soliton complex supported by sixth-order Bessel lattice at $b=5$ and $p=40$. All quantities are plotted in arbitrary dimensionless units.

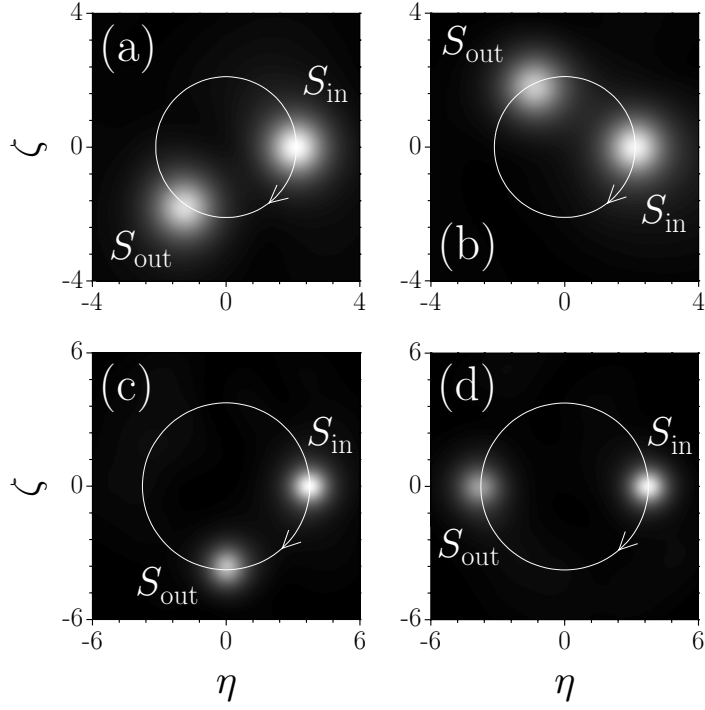

Figure 5. Azimuthal switching of soliton beam to second (a) and fourth (b) main guiding channels of third-order Bessel lattice imprinted in a cubic medium with nonlinearity saturation, for input angles $\alpha_\zeta = 0.49$ and $\alpha_\zeta = 0.626$ at $p = 2$. Energy flow of input soliton $U_{\text{in}} = 8.26$. (c) and (d) show switching to third and sixth channels of sixth-order lattice for $\alpha_\zeta = 0.8$ and $\alpha_\zeta = 0.93$ at $p = 5$. Energy flow of input soliton in (c) and (d) $U_{\text{in}} = 8.61$. Input and output intensity distributions are superimposed onto each other. Arrows show direction of input soliton motion and labels $S_{\text{in}}, S_{\text{out}}$ denote input and output soliton positions. Saturation parameter $S = 0.1$. All quantities are plotted in arbitrary dimensionless units.

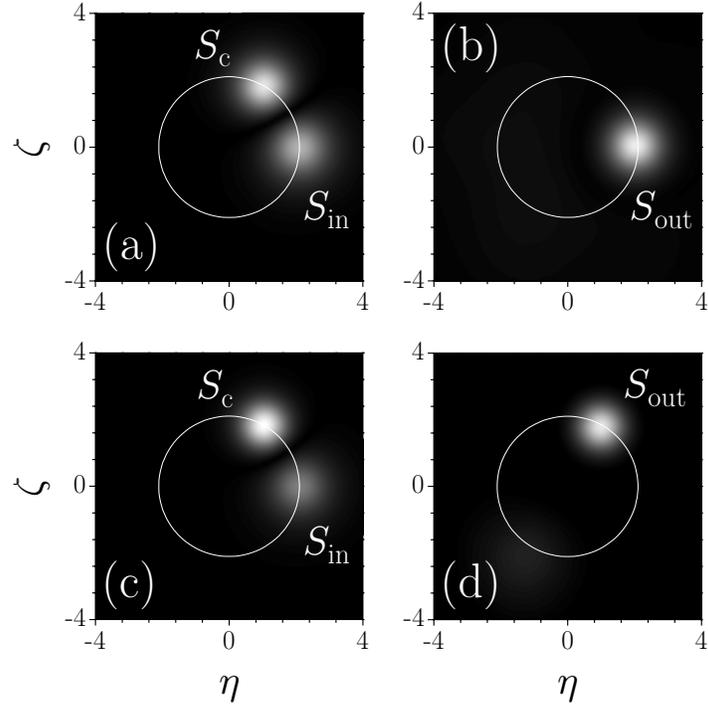

Figure 6. Controllable fusion of a soliton and control soliton beam launched into fifth guiding channel of third-order Bessel lattice imprinted in a cubic medium with nonlinearity saturation. Input (a) and output (b) intensity distributions for control beam energy $U_c = 10.64$. Input (c) and output (d) intensity distributions for $U_c = 20.29$. The input soliton energy is $U_{in} = 8.26$ and it is initially located in the zero channel. Labels $S_{in}, S_{out}, S_c$ denote input, output, and control beam positions. Saturation parameter $S = 0.1$, lattice depth $p = 2$. All quantities are plotted in arbitrary dimensionless units.